\documentclass[graybox]{svmult}

\usepackage{mathptmx}        % selects Times Roman as basic font
\usepackage{helvet}          % selects Helvetica as sans-serif font
\usepackage{courier}         % selects Courier as typewriter font
\usepackage{makeidx}         % allows index generation
\usepackage{graphicx}        % standard LaTeX graphics tool
\usepackage{multicol}        % used for the two-column index
\usepackage[bottom]{footmisc}% places footnotes at page bottom

\usepackage{url}             % handles special characters in e-mail

\makeindex             % used for the subject index
\usepackage[T1]{fontenc}
\usepackage[utf8]{inputenc}
\usepackage{url}
\usepackage{hyperref}
\usepackage[UKenglish]{babel}
\usepackage{amsmath}
\usepackage{amssymb}
\usepackage{amstext}
\usepackage{dsfont}
\usepackage[decimalsymbol=.]{siunitx}
\usepackage{rotating}
\usepackage[sort&compress,square,numbers]{natbib}
\newcommand{\Reynolds}{Re}
\newcommand{\Womersley}{W\!o}
\newcommand{\laplacian}{\Delta}
\renewcommand{\nabla}{\begin{sideways}\begin{sideways}%
                      $\Delta$%
                      \end{sideways}\end{sideways}}
\begin{document}

%\frontmatter%%%%%%%%%%%%%%%%%%%%%%%%%%%%%%%%%%%%%%%%%%%%%%%%%%%%%%

%\include{dedic}
%\include{foreword}
%\include{preface}
%\include{acknow}

%\tableofcontents
%\include{cblist}
%\include{acronym}

\mainmatter%%%%%%%%%%%%%%%%%%%%%%%%%%%%%%%%%%%%%%%%%%%%%%%%%%%%%%%
%\include{part}
%%%%%%%%%%%%%%%%%%%% author.tex %%%%%%%%%%%%%%%%%%%%%%%%%%%%%%%%%%%
%
% sample file for your "contribution" to a contributed volume
%
% Use this file as a template for your own input.
%
%%%%%%%%%%%%%%%% Springer %%%%%%%%%%%%%%%%%%%%%%%%%%%%%%%%%%%
{                                                    % scope of contribution 103
%
%
%
%--- title --------------------------------------------------------------------%
\title*{Turbulent kinetic energy transport in oscillatory pipe flow}
\titlerunning{Turbulent kinetic energy transport in oscillatory pipe flow}
\author{Claus Wagner \& Daniel Feldmann}
\authorrunning{C. Wagner \& D. Feldmann}
\institute{Claus Wagner \& Daniel Feldmann (ORCID: 0000-0002-6585-2875) \at
           German Aerospace Center (DLR),
           Institute of Aerodynamics and Flow Technology,
           SCART, \newline
           Bunsenstr. 10, G\"ottingen, Germany, \email{feldmann@mailbox.org} \newline
          %Published version: https://doi.org/10.1007/978-3-319-14448-1\_31
           Published version: \href{https://doi.org/10.1007/978-3-319-14448-1_31}{DOI: 10.1007/978-3-319-14448-1\_31}
          }
\maketitle
%------------------------------------------------------------------------------%
%
%
%
%
%
%
\section{Introduction}
\label{103_sec:introduction}
Laminar as well as turbulent oscillatory pipe flows occur in many fields of
biomedical science and engineering.
Pulmonary air flow and vascular blood flow are usually laminar,
because shear forces acting on the physiological system ought to be small.
However, frictional losses and shear stresses vary considerably with transition
to turbulence.
This plays an important role in cases of e.g. artificial
respiration or stenosis.
On the other hand, in piston engines and reciprocating thermal/chemical
process devices, turbulent or transitional oscillatory flows affect
mixing properties, and also mass and heat transfer.
In contrast to the extensively investigated statistically steady wall bounded
shear flows, rather little work has been devoted to the onset, amplification and
decay of turbulence in pipe flows driven by an unsteady external force.
Experiments \cite{hino1976, eckmann1991, zhao1996} indicate that
transition to turbulence depends on only one parameter, i.e.
$\Reynolds_{\delta}\sim\Reynolds/\Womersley$ with a critical value of about
$550$, at least for Womersley numbers $\Womersley>7$.
We perform direct numerical simulations (DNS) of oscillatory pipe flows at
several combinations of $\Reynolds$ and $\Womersley$ to extend the validity of
this critical value to higher $\Womersley$.
To better understand the physical mechanisms involved during decay and
amplification of the turbulent flow, we further analyse the turbulent kinetic
energy distribution and its budgets terms.
\section{Numerical approach}
\label{103_sec:numericalApproach}
We consider a Newtonian fluid confined by a straight pipe of diameter $D$ and
length $L$.
The fluid is driven in axial direction ($z$) by the time dependent pressure
gradient
%
%--- pressure gradient --------------------------------------------------------%
\begin{align}
\textstyle
\vec{P}(t) \equiv \left[0, 0, \partial_{z}\langle p \rangle_{\phi}\right]^{\intercal}
 = \left[0, 0,-4\cos\!\!\left( \frac{4 \Womersley^{2}}{ \Reynolds_{\tau}} t \right)\right]^{\intercal}
\text{, where}
\hspace{03.50mm}
p = p^{\prime} + \langle p \rangle_{\phi}
\label{103_eq:pressureGradient}
\end{align}
%------------------------------------------------------------------------------%
%
according to Reynolds' decomposition.
Prime denotes the fluctuating part and angle brackets the mean quantity averaged
over equal oscillation phases.
Normalisation and the set of non-dimensional control parameters is given by the
Womersley number $\Womersley = D/2\sqrt{\omega/\nu}$ and the friction Reynolds
number $\Reynolds_{\tau} = u_{\tau}D/\nu$.
Here, $\omega=2\pi/T$ is the forcing frequency, $\nu$ the kinematic viscosity,
and $u_{\tau}$ the friction velocity of a fully developed statistically steady
turbulent pipe flow.
Thus, the governing equations read
%
%--- conservation laws --------------------------------------------------------%
\begin{align}
\textstyle
\vec{\nabla} \cdot \vec{u} = 0
\hspace{07.50mm}
\text{and}
\hspace{07.50mm}
\partial_{t} \vec{u} +
\left( \vec{u} \cdot \vec{\nabla} \right) \vec{u} +
\vec{\nabla} p^{\prime} -
\frac{1}{\Reynolds_{\tau}} {\boldsymbol\laplacian} \vec{u} = \vec{P(t)}
\label{103_eq:conservationLaws}
\end{align}
%------------------------------------------------------------------------------%
%
with $\vec{u}$ denoting the velocity vector, $\partial_{t}$ being the partial
derivative with respect to time $t$ and $\vec{\nabla}$ and
${\boldsymbol\laplacian}$ being the Nabla operator and the Laplacian,
respectively.
Eqs.~(\ref{103_eq:pressureGradient}) and (\ref{103_eq:conservationLaws}) are
supplemented by periodic boundary conditions (BC) for $\vec{u}$ and $p$ in the
homogeneous directions $z$ and $\varphi$ and no-slip and impermeability BC at
$r=D/2$ in the radial direction.
They are directly solved by means of a fourth order accurate finite volume
method and advanced in time using a second order accurate leapfrog--Euler time
integration scheme.
Further details on the numerical method are given in Feldmann \& Wagner
\cite{feldmann2012} and references therein.
The initial flow field was taken from a well correlated statistically steady
turbulent pipe flow at $\Reynolds_{\tau}=1440$ discussed in
\cite{feldmann2012}.
The criterion $\overline{h}^*<\pi(\Reynolds_{\delta}/\langle k^{*}_{\epsilon}\rangle_{z,\varphi,\phi})^{1/4}$
for the mean grid size leads to $\overline{h}^*= \{6.2; 7.1; 5.7\}$ for cases I
to III based on the maximum turbulent dissipation rates plotted in
fig.~\ref{103_fig:tkeBudgets}.
Since the mean grid spacing varies from the wall to the axis between
$0.6<\overline{h}^{*}<1.0$, $1.3<\overline{h}^{*}<2.2$, and
$0.6<\overline{h}^{*}<1.1$, respectively, we conclude, that the used grids are
sufficiently fine to resolve all relevant length scales.
\section{Computational parameters and flow regimes}
\label{103_sec:parametersRegime}
We focus on DNS results obtained for three combinations of
$\Womersley=\left\{13,26,52\right\}$ and
$\Reynolds_{\tau}=\left\{1440,5760,11520\right\}$, resulting in different peak
Reynolds numbers $\Reynolds = \hat{u}D/\nu$ based on the maximum value of the
respective bulk velocity $\langle\bar{u}(t)\rangle_{\phi}$ within $0 <t\leq T$.
Here, $\phi$ symbolises averaging over $N=14$ equal phases with $t+nT/2$ for
$n=\{n\in\mathbb N:n\leqslant N\}$.
The resulting parameter space in terms of $\Reynolds$ and $\Womersley$ is shown
in fig.~\ref{103_fig:parameterRegime}, where
$\Reynolds_{\delta} = \delta \hat{u}/\nu = \Reynolds/\Womersley\sqrt{1/2}$
is the Reynolds number based on the Stokes layer thickness
$\delta = \sqrt{2\nu/\omega}$.
%
%--- parameter regime ---------------------------------------------------------%
\begin{figure}[t]
\sidecaption
\includegraphics[width=0.60\textwidth]{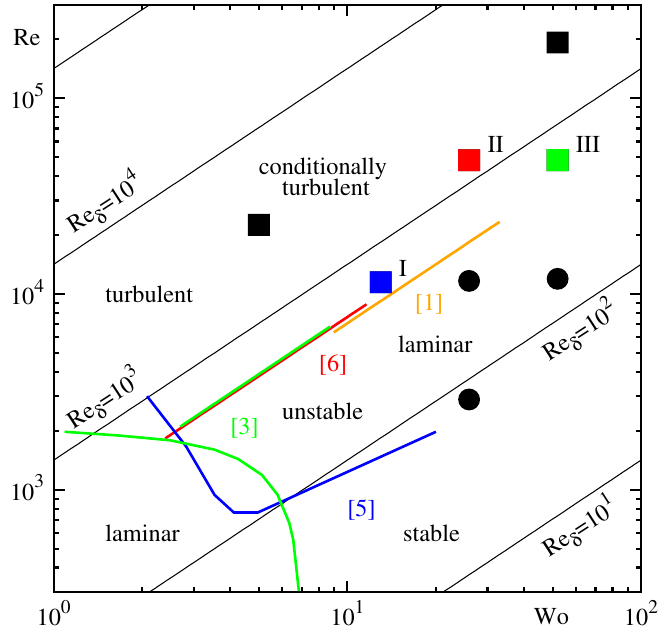}
\caption[Non-dimensional parameters and flow regimes]
        {Non-dimensional parameter space and flow regimes in terms of
         $\Reynolds$ and $\Womersley$. Circles denote laminarised flows and
         squares turbulent flows. Here, we focus on the turbulent
         cases I, II, and  III. Transition from laminar to the turbulent flow
         regime is indicated by the green \cite{hino1976}, red \cite{zhao1996},
         and orange \cite{eckmann1991} lines as obtained experimentally.
         The blue line \cite{trukenmueller2006} separates regions of stability
         according to a quasi-steady stability analysis.}
\label{103_fig:parameterRegime}
\end{figure}
%------------------------------------------------------------------------------%
%
All flows denoted by circles laminarise despite of high $\Reynolds$ up to
$\sim12000$.
The stabilising effect of the oscillatory forcing increases with $\Womersley$,
at least beyond a certain value of about seven, in agreement with findings from
stability analysis \cite{trukenmueller2006} and experimental investigations
\cite{hino1976, zhao1996, eckmann1991}.
\par
Fig.~\ref{103_fig:timeSeries} presents time series of the applied forcing
$\vec{P}(t)$, the predicted mean shear stress at the wall
$\langle\tau^{*}_{w}\rangle_{z,\varphi}$, and the axial velocity component
$u_{z}(r,t)$ close to the wall ($r/D = 0.49$) and near the center line
($r/D=0.01$).
%
%--- time series --------------------------------------------------------------%
\begin{figure}[b]
\sidecaption
\includegraphics[width=0.33\textwidth]{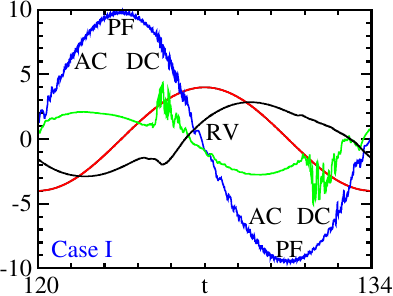}
\includegraphics[width=0.33\textwidth]{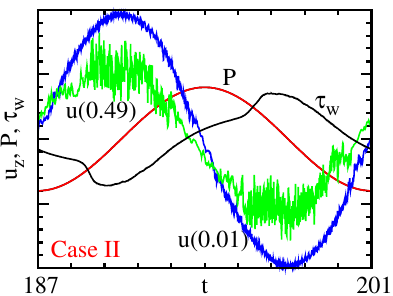}
\includegraphics[width=0.33\textwidth]{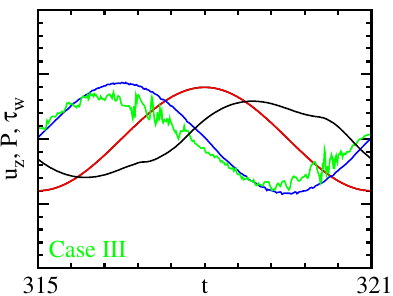}
\caption[Time series]
        {Time series of the axial velocity component $u_{z}(r,t)$ close to the
         wall at $r/D=0.49$ and close to the centre line at $r/D=0.01$, the
         forcing $\vec{P}(t)$, and the mean wall shear stress
         $\langle\tau^{*}_{w}\rangle_{z,\varphi}$.}
\label{103_fig:timeSeries}
\end{figure}
%------------------------------------------------------------------------------%
%
The time series of $u_{z}$ reveal conditionally turbulent flows for all three
parameter combinations, i.e. case I at $\Womersley=13$ and
$\Reynolds=11460$, case II at $\Womersley=26$ and $\Reynolds=48175$,
and case III at $\Womersley=52$ and $\Reynolds=48250$, respectively.
For I and III, i.e. the two slightly supercritical cases both at
$\Reynolds_{\delta}\sim600$, the near wall velocity characteristics are similar.
Fluctuations suddenly grow only at deceleration (DC), while they are
damped again during flow reversal (RV) and the following acceleration (AC)
phase.
These turbulent bursts during DC are also reflected in the
$\tau^{*}_{w}$ history.
However, the most conspicuous difference between I (lower $\Womersley$) and III
(higher $\Womersley$) in this respect is that the AC phase is more stable for
$\Womersley=13$ despite of the similar $\Reynolds_{\delta}$, while
fluctuations in the near wall flow do not completely decay for higher
$\Womersley$.
Contrarily, the $u_{z}$ history at the centre line is completely smooth for
case III, while for case I the core region is characterised by substantial
velocity fluctuations throughout the whole oscillation cycle.
\par
For higher values of $\Reynolds_{\delta}$, i.e. case II, the velocity time
series reveal a completely different behaviour without distinct bursts and
strong velocity fluctuations at the centre line.
The turbulence intensity close to the wall and in the core region rather
increases during AC and decreases during DC analogously to the bulk flow with
laminarisation only during RV.
\section{Turbulent kinetic energy}
\label{103_sec:turbulentKineticEnergy}
To shed light on the mechanisms leading to the different behaviour in decay and
amplification of turbulence in the oscillatory pipe flows discussed above, we
analyse the turbulent kinetic energy $k^{*}$ as well as the production and
dissipation terms of its transport equation, see e.g. \cite{feldmann2012}.
During all oscillation phases, both flows at $\Reynolds_{\delta}\sim\num{600}$
develop a boundary layer with one major characteristic, which is typical for
wall-bounded shear flows.
The turbulent kinetic energy profiles exhibit an obvious maximum very
close to the wall with a very steep decrease towards the wall ($r^{*} = 0$) and
a moderate drop towards the pipe centre line.
This can be seen from the radial $k^{*}$ distribution for I and III, shown in
fig.~\ref{103_fig:tkeBudgets}.
%
%--- tke budgets --------------------------------------------------------------%
\begin{figure}[htbp]
\includegraphics[width=0.33\textwidth]{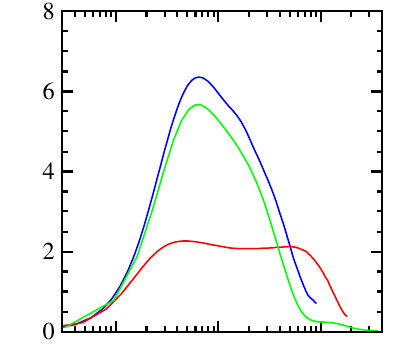}\hfill
\includegraphics[width=0.33\textwidth]{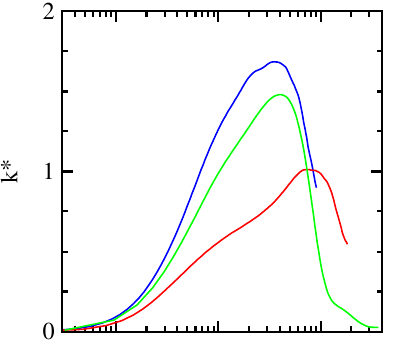}\hfill
\includegraphics[width=0.33\textwidth]{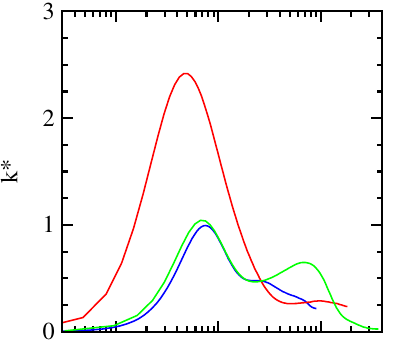}\\
\includegraphics[width=0.33\textwidth]{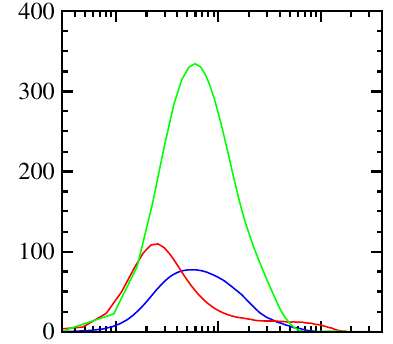}\hfill
\includegraphics[width=0.33\textwidth]{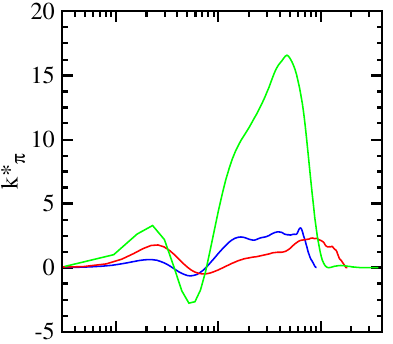}\hfill
\includegraphics[width=0.33\textwidth]{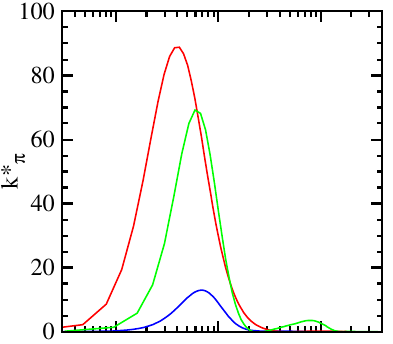}\\
\includegraphics[width=0.33\textwidth]{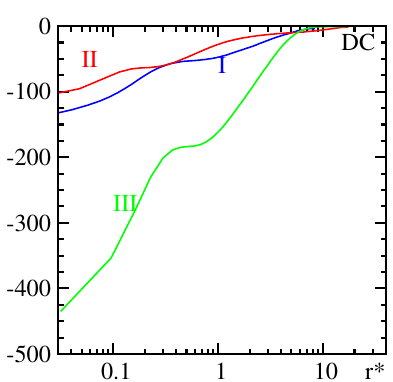}\hfill
\includegraphics[width=0.33\textwidth]{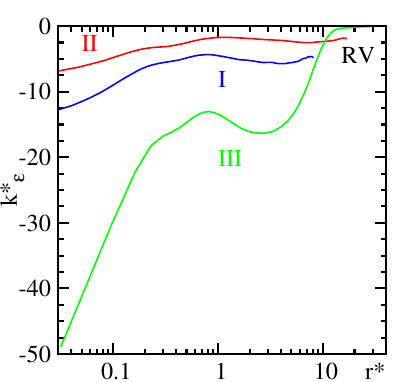}\hfill
\includegraphics[width=0.33\textwidth]{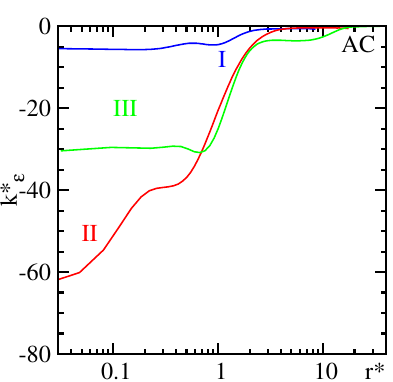}
\caption[Turbulent kinetic energy profiles]
        {Radial distribution of turbulent kinetic energy $k^*$, its
         production term $k^*_{\pi}$, and its turbulent dissipation term
         $k^*_{\epsilon}$ scaled in Stokes layer thicknesses during DC, RV, and
         AC for
         I:   $\Womersley=13$, $\Reynolds_{\delta}=623$;
         II:  $\Womersley=26$, $\Reynolds_{\delta}=1310$; and
         III: $\Womersley=52$, $\Reynolds_{\delta}=656$.}
\label{103_fig:tkeBudgets}
\end{figure}
%------------------------------------------------------------------------------%
%
They reach the same maximum value for similar $\Reynolds_{\delta}$ and thus the
ratio of $\Reynolds$ to $\Womersley$ is the governing parameter defining equally
turbulent oscillatory pipe flows.
However, due to the shorter oscillation phase in case III, i.e. a four times higher
$\Womersley$, the same amount of energy is produced in a shorter period of time,
as reflected by the four times higher production rate $k_{\pi}$ presented in
fig.~\ref{103_fig:tkeBudgets}.
For I and III, the turbulent kinetic energy monotonically decays from DC via RV
even until AC.
The production term $k^{*}_{\pi}$ further confirms, that turbulence is mostly
generated when turbulent near-wall bursts occur during DC, cf.
fig.~\ref{103_fig:timeSeries}.
During RV, more energy is dissipated than produced at a dissipation rate
$k^{*}_{\pi}$ even higher as during AC.
Furthermore, for $\Reynolds_{\delta}\sim600$ the $k^{*}_{\pi}$ profile becomes
negative in a small annular region, where turbulent kinetic energy is
transferred back to the mean flow, see also \cite{feldmann2012}.
Both phenomena in combination account for the rigorous laminarisation during RV.
\par
In contrast, for $\Reynolds_{\delta}=1310$ the turbulent kinetic energy
decreases to the half during RV and increases again during AC to about the same
value as in DC, due to the highest production rate $k^{*}_{\pi}$ during AC.
Case II is also different in such a way that the turbulent kinetic energy is
also rather high in most of the core region during DC, characterised by a flat
broad $k^*$ profile only showing a vague near-wall maximum.
While during AC the $k^*$ distribution clearly exhibits the above described
typical shear flow profile.
The same typical shear flow behaviour is reflected by the $k^{*}_{\epsilon}$
distribution with a maximum dissipation at the wall and a distinct plateau
during all oscillation phases for $\Reynolds_{\delta}=1310$.
The slightly supercritical cases I and III at $\Reynolds_{\delta}\sim600$, on
the other hand, develop a second local dissipation maximum, which becomes more
pronounced in the $k^{*}_{\epsilon}$ profiles with ongoing laminarisation
during RV and AC.
\par
The wall distance $r^{*} = \left(1/2 - r \right) D/\delta$ of the $k^{*}$ maxima
are also the same for similar $\Reynolds_{\delta}$, when the length is scaled in
Stokes layer thicknesses $\delta$.
During DC and AC, when the bulk flow is high, the $k^{*}$ maximum occurs closer to
the wall for higher $\Reynolds_{\delta}$.
Vice versa, the energy maximum occurs closer to the wall for lower
$\Reynolds_{\delta}$ during RV, when viscous effects dominate.
Thus, the ratio of $\Reynolds$ to $\Womersley$ defines the thickness of
the annular region in terms of $\delta$, in which the oscillatory pipe flow
exhibits turbulent features, even though the thickness of this turbulent
boundary layer is strongly phase dependent.
Also the budget terms shown in fig.~\ref{103_fig:tkeBudgets}, reveal that the
wall distance of all the characteristics in the profiles, i.e. maxima,
inflection points, plateaus and so on, scale with $\Reynolds_{\delta}$.
However, for decreasing $\Womersley$ per definition the Stokes layer becomes
large compared to the pipe radius and thus the geometrical constraint of the
pipe wall gains importance.
In case I turbulent near-wall structures evolve during DC and RV, penetrate
farther towards the centre line, and thus span almost the whole core region,
cf. fig.~\ref{103_fig:timeSeries}.
As a result, the turbulent kinetic energy is rather high over most parts of the
pipe radius and, more important, does not vanish at the pipe axis for lower
$\Womersley$.
Whereas, in case III at higher $\Womersley$ but equal $\Reynolds_{\delta}$
the $k^{*}$ distribution reflect a high turbulent kinetic energy for $r^*<9$ and
an almost laminar flow over the second half of the pipe radius.
As indicated before by the velocity time series shown in
fig.~\ref{103_fig:timeSeries}, the turbulence is confined to a smaller (in terms
of $r$) annular region close to the wall, while the flow remains very smooth in
the whole core region for the highest $\Womersley$.
%
% \par
%
The contribution of all other transport terms, i.e. the viscous, turbulent, and
pressure diffusion as well as the pressure strain, to the overall budget is much
lower.
In principle, these terms reflect the typical shear flow mechanisms, which are
simply damped and amplified by the oscillatory forcing, and thus for the sake of
brevity not further discussed here.
\section{Conclusions}
\label{103_sec:conclusions}
Decay, amplification, and redistribution of turbulent kinetic energy in
oscillatory pipe flow were studied by means of DNS for various combinations of
$\Reynolds$ and $\Womersley$.
We found, that oscillatory flows at $\Reynolds_{\delta}<550$ relaminarise when
started from a fully developed turbulent flow field despite of high $\Reynolds$.
In very good agreement with experiments \cite{hino1976, eckmann1991, zhao1996},
we confirm oscillatory flows at $\Reynolds_{\delta}>550$ to be conditionally
turbulent.
However, we contradict \cite{eckmann1991} who stated that core flow remains
stable for $\Reynolds_{\delta} < 1310$.
Our DNS results extend the validity of this experimentally determined
critical value up to $\Womersley=52$ in the $\Reynolds$-$\Womersley$-space.
Nevertheless, from the analysed turbulent kinetic energy distributions we
conclude that decay and amplification of turbulence in oscillatory pipe
flows rather depend on the combination of $\Reynolds$ and $\Womersley$ then on
its ratio ($\Reynolds_{\delta}$) alone.
This is a significant difference to the case for the oscillating boundary layer
over a flat plate, which has been extensively studied by Spalart \& Baldwin
\cite{spalart1989} using DNS.
Even if experiments have shown that, the transition to turbulence can be
characterised by only using $\Reynolds_{\delta}$, at least for
$\Womersley>7$, our study revealed that for oscillatory pipe flow the
additional geometrical constraint considerably affects the decay and
amplification of turbulence in the core flow.

}                          % end of command definition scope of contribution 103
%------------------------------------------------------------------------------%
%
%
%

%

\backmatter%%%%%%%%%%%%%%%%%%%%%%%%%%%%%%%%%%%%%%%%%%%%%%%%%%%%%%%
\appendix
%\include{appendix}
%\include{glossary}
%\printindex

%%%%%%%%%%%%%%%%%%%%%%%%%%%%%%%%%%%%%%%%%%%%%%%%%%%%%%%%%%%%%%%%%%%%%%

\end{document}